\documentstyle[12pt]{article}
\newcommand{\uq}{U_q(C(n+1))}
\newcommand{\lam}{\lambda}
\newcommand{\Lam}{\Lambda}
\newcommand{\x}{\otimes}
\newcommand{\e}{\epsilon}
\newcommand{\del}{\delta}
\newcommand{\ba}{\begin{eqnarray}}
\newcommand{\na}{\end{eqnarray}}
\newcommand{\ban}{\begin{eqnarray*}}
\newcommand{\nan}{\end{eqnarray*}}
\newcommand{\dimq}{dim_{{\bf C}[q, q^{-1}]}}
\newcommand{\dimc}{dim_{\bf C}}
\newtheorem{lemma}{Lemma}
\newtheorem{theorem}{Theorem}
\newtheorem{proposition}{Proposition}

\begin{document}
\title{Finite Dimensional Representations of $\uq$ at Arbitrary $q$}
\author{R. B. Zhang\\
        Department of Theoretical Physics\\
        Research School of Physical Sciences and Engineering\\
        The Australian National University\\
        Canberra, ACT 0200, Australia }
\date{ }
\maketitle

\vspace{3cm}
\noindent
A method is developed to construct irreducible representations(irreps)
of the quantum supergroup $\uq$ in a systematic fashion. It is shown
that every finite dimensional irrep of this quantum supergroup at
generic $q$ is a deformation of a finite dimensional irrep of
its underlying Lie superalgebra $C(n+1)$, and is essentially uniquely
characterized by a highest weight. The character of the irrep is given.
When $q$ is a root of unity, all irreps of $\uq$ are finite dimensional;
multiply atypical highest weight irreps and (semi)cyclic irreps also
exist. As examples, all the highest weight and (semi)cyclic irreps of
$U_q(C(2))$ are thoroughly studied.

\pagebreak
\section{Introduction}
This is the third of a series of papers systematically developing
the representation theory of the quantum supergroups\cite{supergroups}
associated with the basic classical Lie superalgebras\cite{kac}.
The first paper\cite{I} studied the structures of the finite dimensional
representations of the quantum supergroup $U_q(gl(m|n))$ at
arbitrary $q$ (The finite dimensional irreps of $U_q(gl(m|1))$ have
been explicitly constructed in \cite{palev}.), while the second
one\cite{II} treated the representation theory of $U_q(B(0, n))$.
It is the aim of the present paper to study $\uq$.

Vigorous study of the theory of Lie superalgebras began in the 1970s
(for reviews see \cite{kac}), largely motivated by the discovery
of supersymmetry in theoretical physics. It was clear from the very
beginning that  although some properties  of ordinary Lie algebras
are shared by their  $\bf{Z}_2$ - graded counterparts,
the Lie superalgebras are  by no means straightforward generalizations
of ordinary Lie algebras; in particular, their representation theory
is drastically different from that of the latter.

Recall that the Weyl groups are of paramount importance in
the study of the finite dimensional irreps of the Lie algebras:
they enable one to compute the characters, which embody all the
information about the weight spaces and dimensions etc of the irreps.
Also, the finite dimensional representations of Lie algebras are
completely reducible. This fact
makes it possible to understand the structures of all finite
dimensional representations.

However, it is not possible in general(except for $osp(1|2n)$)
to introduce a Weyl group for a
Lie superalgebra, which is not simply the Weyl group of the
maximal even subalgebra.  The so called Weyl groups of
Lie superalgebras embody little useful information
about the odd generators, thus not allowing the determination
of the structures of  irreps.
Also, finite dimensional representations of Lie superalgebras
are not completely reducible.
These facts make the representation theory of Lie superalgebras
an extremely difficult subject to study.

Quantum supergroups are one parameter deformations of the
universal enveloping algebras
of basic classical Lie superalgebras, which were only introduced
couple a years ago\cite{supergroups}, and have been intensively
studied ever since.
It has become clear that the quantum supergroups are of great
importance to the study
of integrable lattice models in statistical mechanics and knot
theory, and also have
deep connections with a variety of other fields in theoretical physics.
In all the applications of quantum supergroups,  their finite
dimensional representations
play a central role. However,  our understanding of such
representations is very incomplete.

It is by now well known that the representation theory of ordinary
quantum groups at generic $q$ is very much the same as that of
the corresponding Lie algebras\cite{lusztig}.
Lusztig and Rosso proved that each
finite dimensional irrep of a quantum group is a deformation
of an irrep of the underlying
Lie algebra, and all finite dimensional representations are
completely reducible.
Rosso's proof made essential use of the properties of the Weyl
groups of the Lie algebras,
while the main ideas of Lusztig's proof are as follows:
In the $q\rightarrow 1$ limit, an
integrable highest weight irrep $\pi$ of a quantum group $U_q(g)$ reduces to
an indecomposible representation $\tilde\pi$ of its underlying
Lie algebra $g$. As integrable representations of ordinary Lie
algebras are completely reducible, $\tilde\pi$, being
indecomposible, must be an irreducible representation of $g$.

Obviously none of these proofs can generalize to quantum supergroups, as
the basic ingredients, namely, Weyl groups and complete
reducibility of integrable representations, are lacking in this  case.
In view of Lusztig's work, it even appears possible intuitively
that a finite dimensional irrep of a quantum supergroup at generic
$q$ may reduce to an indecomposible
but reducible representation of the underlying Lie superalgebra
in the limit $q\rightarrow 1$.  Fortunately it turned out not
to be the case, at least for $U_q(gl(m|n))$ and $U_q(osp(1|2n))$,
as shown in \cite{I} and \cite{II}.

One of the main results of the present paper is the proof that
every finite dimensional irrep of the quantum supergroup $\uq$
at generic $q$ reduces to an irrep of the underlying
Lie superalgebra $C(n+1)$, and the two irreps have the same weight space
decomposition. This result enables us to gain a rather thorough
understanding of the structures of finite dimensional irreps of $\uq$,
in particular, to write down their character
formula, as $C(n+1)$ happens to be one of the very few Lie
superalgebras having a well developed representation theory\cite{jeugt}.

When $q$ is a root of unity, we will develop a method allowing
us to  construct $\uq$ irreps in a systematic fashion.
The representation theory in this case changes dramatically,
in particular, all irreps are finite dimensional, (semi)cyclic
irreps and multiply atypical irreps appear.

The arrangement of the paper is as follows.
In section 2, we prove a generalized BPW theorem for $\uq$. In
section 3 we generalize Kac' induced module construction for
Lie superalgebras to this quantum supergroup at arbitrary $q$,
and also thoroughly investigate the structures
of the finite dimensional irreps at generic $q$. In section 4
we explicitly construct all the irreps of $U_q(C(2))$ using
the general theory developed in the earlier sections.

\section{BPW theorem for $\uq$}
This section studies the structure of the ${\bf Z}_2$ graded
Hopf algebra $\uq$. In particular, a generalized BPW theorem
for this  quantum supergroup will be proved,
and an explicit basis for it will also be constructed.
Results of this section will be repeatedly applied throughout
the paper.

\subsection{Definition of $\uq$}
Let us begin by defining the quantum supergroup $\uq$.
Recall that Lie superalgebras admit different simple root systems,
which can not be mapped to one another by the Weyl groups of
their maximal even subalgebras.
As quantization treats the Cartan and simple generators differently
from the rest, it is not clear whether the quantum supergroups
obtained by quantizing the same Lie superalgebra but using
different simple root systems are algebraically
equivalent(It is not difficult to convence oneself by examining
simple examples that the
resultant quantum supergroups are co - algebraically inequivalent).
However, we will not be concerned with this problem here, but merely
take $\uq$ as the quantization of the universal enveloping algebra of
the type I superalgebra $C(n+1)$ with the commonly used
simple root system, namely, the one with a unique odd simple root.

To describe this simple root system, we introduce the $(n+1)$
- dimensional Minkowski space $H^*$ with a basis
$\{\delta_i | i=0, 1, 2, ...,n\}$ and the bilinear form
$(\ ,\ ): H^*\times H^*$ $\rightarrow C$ defined by
\ban
(\delta_i,\ \delta_j)=-(-1)^{\del_{0i}} & \forall i, j.
\nan
Then, following Kac,  the simple roots of $C(n+1)$ can be expressed as
\ban
\alpha_i&=&\delta_i-\delta_{i+1}, \ \ i=0, 2, ..., n-1, \\
\alpha_n&=&2\delta_n,
\nan
where $\alpha_0$ is the unique odd simple root.
A convenient version of the  Cartan matrix
$A=(a_{ij})_{i,j=0}^n$   for $C(n+1)$ is
\ban
a_{ij}&=&2(\alpha_i, \alpha_j)/(\alpha_i, \alpha_i), \ \   \forall i>0,\\
a_{0j}&=&(\alpha_0, \alpha_j).
\nan
We denote by $\Delta_0^+$ and $\Delta_1^+$
the set of the even positive roots and that of the odd positive
roots of $C(n+1)$ respectively. Then
\ban
\Delta_0^+&=&\{ \delta_i-\delta_j, \
\delta_i+\delta_j, \ 2\delta_i | 0<i<j\}, \\
\Delta_1^+&=&\{ \delta_0\pm \delta_i| i>0\}.
\nan
For later use, we also define
\ban
\rho_{\theta}&=&{1\over 2}\sum_{\alpha\in \Delta^+_{\theta}} \alpha,
\ \ \  \ \theta=1,2, \\
\rho&=&\rho_0-\rho_1.
\nan

Let $q$ be an indeterminant, and define
$$q_i =\left\{ \begin{array}{ll}
               q^{(\alpha_i, \alpha_i)/2}, & i>0, \\
               q,                          & i=0.
               \end{array}
        \right.  $$
The quantum supergroup $\uq$ is the unital ${\bf Z}_2$ -
graded algebra on the field ${\bf C}[q, q^{-1}]$, which is  generated by
$\{ e_i,\ f_i \ K_i^{\pm}|\ i = 0, 1,..., n\}$ with the relations
\ban
K_i^{\pm 1}K_j^{\pm 1}&=&K_j^{\pm 1}K_i^{\pm 1},\\
K_i K_i^{-1}&=&1, \\
K_i e_jK_i^{-1}&=& q_i^{a_{ij}}e_{j}, \\
K_if_jK_i^{-1}&=&q_i^{-a_{ij}}f_{j},\\
{[}e_i, f_j{\rbrace}&= &\del_{ij}(K_i-K_i^{-1})/(q_i-q_i^{-1}),
\nan
\ba
(e_0)^2=(f_0)^2=0,\nonumber \\
\sum_{\mu=0}^{1-a_{ij}}(-1)^\mu
\left[\begin{array}{c}
       1-a_{ij}\\
      \mu
      \end{array}
\right]_{q_{i}}
e_i^{1-a_{ij}-\mu }e_je_i^\mu =0, &i\ne 0, \nonumber \\
\sum_{\mu=0}^{1-a_{ij}}(-1)^\mu
\left[\begin{array}{c}
       1-a_{ij}\\
      \mu
      \end{array}
\right]_{q_{i}}
f_i^{1-a_{ij}-\mu }f_jf_i^\mu =0, & i\ne 0, \label{def}
\na
where
\ban
\left[\begin{array}{l}
       m\\
       n
      \end{array}
\right]_q
&=&{{[m]_q!}\over{[n]_q![m-n]_q!}}, \ \ \ m\ge n,  \\
{[k]}_q!&=&\left\{\begin{array}{ll}
             \prod_{i=1}^k{{q^i-q^{-i}}\over{q-q^{-1}}}, &k>0, \\
              1,             &k=0.
             \end{array}
        \right.
\nan
In (\ref{def}), $[x, y{\rbrace}=xy-(-1)^{[x][y]}yx$,
with the gradation indices $[x]$ and $[y]$ defined by
$$[K_{i}]=0, \ \ \ \forall i, \ \ \ \ \ \
{[}e_{i}]=[f_{i}]=\left \{\begin{array}{ll}
                           0,&i>0\\
                           1,&i=0,
                          \end{array}
                          \right. $$
and for  any $u, v$ which are monomials in
$e_{i}, f_{i}, K_{i}^{\pm 1}, \ i=0, 1,...,n,\ $
$[uv]{\equiv}[u]+[v](mod 2)$. This makes  $\uq$  a
${\bf Z}_{2}$--graded algebra with $\uq=U_{0}{\oplus}U_{1}\ $,
where $U_{0}=\lbrace u{\in}U_{q}(g)|[u]=0\rbrace\ $,
$U_{1}=\lbrace u{\in}U_{q}(g)|[u]=1\rbrace$.
We will call elements of $U_0$ even and those of $U_1$ odd.
We also  associate with their product $u v$ an element of
$H^*$, $wt(uv)=wt(u)+wt(v)$, called the weight,
such that $wt(e_i)=-wt(f_i)=\alpha_i$,
$wt(K_i^{\pm 1})=0$. If $w\in\uq$ is a linear combination
of monomials of the same weight $\mu\in H^*$, we define $wt(w)=\mu$.

The quantum supergroup $\uq$ has the structures of a ${\bf Z}_2$
graded Hopf algebra with
invertible antipode. One consistent  co - multiplication reads,
\ban
\Delta(K_i^{\pm 1})&=&K_i^{\pm 1}\x K_i^{\pm 1}\\
\Delta(e_i)&=&e_i\x 1 + K_i\x e_i,\\
\Delta(f_i)&=&f_i\x K_i^{-1} + 1\x f_i;
\nan
and the corresponding co - unit $\e$ and antipode $S$ are
respectively given by
\ban \e(e_i)&=&\e(f_i)=0, \\
\e(K_i)&=&\e(K_i^{-1})=1, \\
S(e_i)&=&- K_i^{-1}e_i, \\
S(f_i)&=&-f_iK_i, \\
S(K_i^{\pm 1}) &=& K_i^{\mp 1}, \  \  \forall i.
\nan

Note that $\{e_i,\ f_i \ K_i^{\pm}|\ i = 1, 2, ..., n\}$
generate a subalgebra $U_q(sp(2n))\subset \uq$.
Together with $\{K_0^{\pm 1}\}$, they  generate
$U_q(sp(2n)\oplus u(1))$
which we will refer to as the maximal even quantum subgroup of $\uq$.

\subsection{$\uq$ at generic $q$}
In order to study the structures of $\uq$, we introduce the
${\bf Z}_2$ graded automorphism
\ban
\varpi(e_i)=f_i,& \varpi(f_i)=e_i, &\varpi(K_i)=K_i,\\
\varpi(q)=q^{-1},&\varpi(c)=c^*, &c\in{\bf C},
\nan
and the anti - automorphism
\ban
\omega(e_i)=f_i, & \omega(f_i)=e_i, & \omega(K_i)=K_i^{-1}, \\
\omega(q)=q^{-1},&  \omega(c)=c^*, &c\in{\bf C},
\nan
which are also required to satisfy
$\varpi(u v)=(-1)^{[u][v]}\varpi(u)\varpi(v)$,
$\omega(u v)= \omega(v)\omega(u)$,
for any homogeneous elements $u, v \in \uq$, and generalize
to all elements of $\uq$ through linearity.

Define the maps  $T_i:\uq\rightarrow \uq$,  $\ i=1, 2, ..., n$, by
\ban
T_i(K_j)&=&K_j K_i^{-a_{ij}}, \ \ \forall j,\\
T_i(e_i)&=&-f_i K_i,\\
T_i(f_i)&=&-K_i^{-1}e_i,\\
T_i(e_j)&=&\sum_{t=0}^{-a_{ij}}(-1)^{t-a_{ij}}q_i^{-t}
{ {(e_i)^{-a_{ij}-t}e_j(e_i)^t}
\over {[-a_{ij}-t]_{q_i}! [t]_{q_i}!}},\\
T_i(f_j)&=&\sum_{t=0}^{-a_{ij}}(-1)^{t}q_i^{-a_{ij}-t}
{{(f_i)^{-a_{ij}-t}f_j (f_i)^t} \over {[-a_{ij}-t]_{q_i}![t]_{q_i}!}},
\ \ \forall j\ne i. \\
\nan
Then
\begin{lemma}
The $T_i$, $i=1,2,.., n$, define algebra automorphisms of $\uq$,
thus generating a group
which will be denoted by $\widehat{W}$. They also satisfy
\ba
T_i \omega &=& \omega T_i, \nonumber\\
T_i^{-1} &=& \varpi T_i \varpi^{-1}. \label{T}
\na
\end{lemma}
{\em Proof}: Restricted to the maximal even quantum subgroup
$U_q(sp(2n)\oplus u(1))$, the $T_i$'s  coincide with the Lusztig
automorphisms\cite{lusztig2} of this quantum group.  Thus we only need
to show that $T_1$ preserves the relations in (\ref{def}) involving
$e_0$ and $f_0$, in  order to prove that $T_i$'s are algebra
homomorphisms of $\uq$, since $T_1$ is the
only map amongst the $T_i$'s which acts nontrivially on $e_0$
and $f_0$. Consider, say, $\{T_1(e_0), T_1(f_0)\}$ when $n>1$. Now,
\ban
T_1(e_0)&=&-e_1e_0+q e_0 e_1, \\
T_1(f_0)&=&- f_0 f_1+q^{-1}f_1 f_0.
\nan
Simple calculations lead to
\ban
\{T_1(e_0), T_1(f_0)\}
&=&{ {K_0K_1-K_0^{-1} K_1^{-1}}\over {q-q^{-1}}}\\
&=&T_1({ {K_0-K_0^{-1}}\over {q-q^{-1}}}).
\nan
The other relations can be checked similarly.
Equation (\ref{T}) can be proved by explicitly working out
the actions of the maps
involved on the simple and Cartan generator of $\uq$.

The maximal even quantum subgroup
$U_q(sp(2n)\oplus u(1))$
admits the following decomposition
\ban
U_q(sp(2n)\oplus u(1))=B_-B_0B_+,
\nan
where $B_ +$ is generated by $\{e_i|i>0\}$, $B_-$ by $\{ f_i|i>0\}$,
and $B_0$ by $\{K_i^{\pm 1}, | i=0,1,..,n\}$.
A basis for $B_0$ is given by
$\{ K^{({\hat r})}H^{(\hat s)} |{\hat r}, {\hat s} \in {\bf Z}^{n+1}_+\}$,
with ${\hat r}=(r_0, r_1, ..., r_n)$,
$K^{({\hat r})} = \prod_{i=0}^n K_i^{r_i}$,
$H^{(\hat r)}=
\prod_{i=0}^n\left( {{K_i - K_i^{-1}}\over {q_i -q_i^{-1}}}\right )^{r_i}$.

Follwing \cite{concini}, we construct bases for $B_+$ and $B_-$
by considering the maximal element $T$ of $\widehat W$,
a reduced expression for which reads
\ban
T&=&T_{i_1}T_{i_2}...T_{i_{n^2}}\\
 &=&(T_1T_2...T_{n-1}T_{n}T_{n-1}...T_{2}T_{1})(T_2...T_{n-1}T_{n}T_{n-1}
...T_{2})...  (T_{n-1}T_nT_{n-1})T_{n}.
\nan
We define
\ban
E_{\beta_1}&=&e_1, \\
F_{\beta_1}&=&f_1, \\
E_{\beta_t}&=&T_{i_1}T_{i_2}...T_{i_{t-1}}(e_{i_t}), \\
F_{\beta_t}&=&T_{i_1}T_{i_2}...T_{i_{t-1}}(f_{i_t}),\ \ t=1,2,...,n^2,
\nan
where $\beta_t\in \Delta_0^+$, and clearly $F_{\beta_t}=\omega(E_{\beta_t})$.
Also observe the following important facts\cite{concini}:
$E_{\beta_t}\in B_+$, $F_{\beta_t}\in B_-$, and
\ba
\{E^{(\hat k)}=(E_{\beta_1})^{k_1}(E_{\beta_2})^{k_2}...
(E_{\beta_{n^2}})^{k_{n^2}}
| {\hat k}\in {\bf Z}_+^{n^2}\},\nonumber \\
\{F^{(\hat k)}=(F_{\beta_{n^2}})^{k_{n^2}}(F_{\beta_{n^2-1}})^{k_{n^2-1}}
...(F_{1})^{k_{1}} | {\hat k}\in {\bf Z}_+^{n^2}\},   \label{b}
\na
form bases for $B_+$ and $B_-$ respectively.

To study the odd elements of $\uq$, we define
\ba
\psi_1&=&e_0,  \nonumber \\
\psi_{i+1}&=&\psi_{i}e_i-q e_i\psi_{i}, \ \  1\le i<n, \nonumber \\
\psi_{- n}&=&\psi_n e_n - q^2 e_n \psi_n,  \nonumber \\
\psi_{- i}&=&\psi_{-i-1}e_i-q e_i\psi_{-i-1}, \ \ \ 0<i<n, \nonumber \\
\phi_{\pm i}&=&\omega(\psi_{\pm i}),  \ \ \ i=1,2,..., n.
\label{odd}
\na
They have the following properties:
\begin{lemma}\label{psipsi}
\ba
\psi_{\pm i} \psi_{\pm j} +q^{\pm 1}\psi_{\pm j} \psi_{\pm i}&=&0,
\ \  i\le j, \nonumber \\
\psi_i\psi_{- j} + q \psi_{- j}\psi_{i}&=&0, \ \ \ \forall i\ne j,  \nonumber
\\
\psi_n\psi_{- n}+q^2\psi_{- n}\psi_n&=&0,   \nonumber \\
\psi_{-i-1}\psi_{i+1}+\psi_{i+1} \psi_{-i-1}
+q\psi_{- i}\psi_i+q^{-1}\psi_i\psi_{- i} &=&0, \ \ \  i<n,
\na
and similar relations for $\phi_i$ and $\phi_{- i}$;
\end{lemma}
\begin{lemma}\label{epsi}
\ba
\psi_j e_i - q^{(\alpha_i, \delta_0 -\delta_j)}e_i\psi_j
&=&\delta_{i j}\psi_{i+1},
\ \ \  \forall i, j, \nonumber \\
\psi_{-j} e_i - q^{(\alpha_i, \delta_0 +\delta_j)}e_i\psi_{-j}
&=&\delta_{i+1, j}\psi_{-i+1},
\ \ \  i>1, \nonumber \\
{[}\psi_{i+1}, f_j\}&=&\delta_{ij}\psi_iK_i q_i^{-1},  \nonumber \\
{[}\psi_{- i}, f_j\}&=&-\delta_{ij}\psi_{-i-1} K_i q_i^{-1},
\na
and similar relations for $\phi_{\pm i}$, where $\psi_{n+1}$ and $\phi_{n+1}$
are understood as $\psi_{-n}$ and $\phi_{-n}$ respectively;
\end{lemma}
\begin{lemma}\label{psiphi}
\ba
\{\psi_{\pm i}, \phi_{\pm i}\}&=&
{ {\Pi_{\pm i} -\Pi_{\pm i}^{-1}}\over{q-q^{-1}} }, \\
\{\psi_{\mu}, \phi_{\nu}\}&=&
      \left\{\begin{array}{ll}
      \sum_{\hat{k}, \hat{r}} c_{\hat{k}, \hat{r}}K^{(\hat r)}
      E^{(\hat k)}, &\mu -\nu>0,\\
      \sum_{\hat{k}, \hat{r}} \tilde c_{\hat{k}, \hat{r}}F^{(\hat k)}
      K^{(\hat r)}, &\mu -\nu<0,
     \end{array}
     \right.   \label{anti}
\na
where
\ban
\Pi_i&=&\prod_{k=0}^{i-1}K_k, \\
\Pi_{-i}&=&\Pi_n\prod_{k=i}^{n}K_k.
\nan
In (\ref{anti}), $\mu, \nu = {\pm 1}, {\pm 2}, ..., {\pm n}$;
$c_{\hat{k}, \hat{r}}\in {\bf C}[q, q^{-1}]$
may be nonvanishing only when $wt(E^{(\hat k)})=$
$sign(\nu)\delta_{|\nu|}-sign(\mu)\delta_{|\mu|}$,
and similarly for $\tilde{c}_{\hat{k}, \hat{r}}\in {\bf C}[q, q^{-1}]$.
\end{lemma}

Define
\ban
\bar{\Gamma}^{(\hat\theta)}&=&
 (\psi_1)^{\theta_1} (\psi_2)^{\theta_2}... (\psi_n)^{\theta_n}
(\psi_{-n})^{\theta_{-n}}(\psi_{-n+1})^{\theta_{-n+1}}...
(\psi_{-1})^{\theta_{-1}}, \ \ \ \theta_{\pm i}=0, 1,  \\
\Gamma^{(\hat\theta)}&=&
(\phi_{-1})^{\theta_{-1}}(\phi_{-2})^{\theta_{-2}}...(\phi_{-n})^{\theta_{-n}}
(\phi_n)^{\theta_n}(\phi_{n-1})^{\theta_{n-1}}...(\phi_1)^{\theta_1},
\ \ \  \theta_{\pm i}=0, 1, \\
\bar{\Gamma}&=&\bar{\Gamma}^{(\theta_{\pm i}=1, \forall i)},\ \\
\Gamma&=&\Gamma^{(\theta_{\pm i}=1, \forall i)}.
\nan
\begin{lemma}\label{Psi}
1). Any product of $\psi$'s(resp. $\phi$'s) can be expressed
as a linear combination of $\bar\Gamma^{(\hat\theta)}$ (resp.
$\Gamma^{(\hat\theta)}$), $\hat\theta\in{\bf Z}_2^{2n}$;\\
2). $\bar\Gamma^{(\hat\theta)}$ (resp. $\Gamma^{(\hat\theta)}$),
$\hat\theta\in{\bf Z}_2^{2n}$ are linearly independent over
${\bf C}[q, q^{-1}]$.
\end{lemma}
{\em Proof}: Part 1) is a direct consequence of Lemma \ref{psipsi}.
To prove 2), we note that any nontrivial relation of the form
$\sum_{\hat\theta}c_{\hat\theta}
\bar\Gamma^{(\hat\theta)}=0$ would lead to $\bar\Gamma \equiv 0$
in $\uq$. Then it would follow that in any linear representation
of $\uq$, the $z$ defined by
\ba
               z&=&\bar{\Gamma}\Gamma,  \label{z}
\na
vanishes identically. But it is easy to construct representations
in which $z$ is nonzero.

For later use, we define the following vector spaces
\ban
\Psi=\bigoplus_{\hat\theta}{\bf C}[q, q^{-1}]\bar{\Gamma}^{(\hat\theta)}, &
\Phi=\bigoplus_{\hat\theta}{\bf C}[q, q^{-1}]\Gamma^{(\hat\theta)}.
\nan
Direct computations can easily establish
\begin{lemma}\label{Gamma}
Let $a\in U_q(sp(2n)\oplus u(1))$, and $b\in U_q(sp(2n))\subset\uq$. Then
\ba
[b, \Gamma]&=&0, \nonumber \\
{[}b, \bar \Gamma]&=&0,\nonumber\\
{[ a, z]}&=&0.
\na
\end{lemma}
{\em Proof}: The first two equations follow from Lemma \ref{epsi}.
They also lead to the last equation.

Now we have the following generalized BPW theorem for the quantum
supergroup $\uq$
\begin{theorem}\label{bpw}
Let $U_+$ (resp. $U_-$)$\in\uq$ be the subalgebra generated by
$\{e_i | i=0,1,...,n\}$ (resp. $\{f_i | i=0,1,...,n\}$). Then\\
1). $\uq$ admits the triangular decomposition
\ba
\uq=U_-B_0U_+;
\na
or more precisely, the multiplication of $\uq$ gives rise to the
${\bf C}[q, q^{-1}]$ module  isomorphism
\ban
U_-\x B_0\x U_+\rightarrow \uq;
\nan
2). $U_+$ and $U_-$ respectively have the bases
\ba
\{E^{(\hat k)}\bar{\Gamma}^{(\hat\theta)} |
{\hat k}\in {\bf Z}_+^{n^2},  {\hat\theta}\in {\bf Z}_2^{2n}\},&
\{\Gamma^{(\hat\theta)}F^{(\hat k)} |
{\hat k}\in {\bf Z}_+^{n^2},  {\hat\theta}\in {\bf Z}_2^{2n}\};
\na
3). The following elements form a basis for $\uq$
\ba
\{\Gamma^{(\hat\theta)}F^{(\hat k)}K^{(\hat r)}H^{(\hat s)}
E^{(\hat l)}\bar{\Gamma}^{(\hat\theta')}
| {\hat k},\hat l\in {\bf Z}_+^{n^2},
\hat r,\hat s\in {\bf Z}_+^{n+1},
{\hat\theta},\hat\theta'\in {\bf Z}_2^{2n}\}.
\na
\end{theorem}
{\em Proof}: 1) is a consequence of the defining relations of
$\uq$. $U_+$ is spanned by  $\{E^{(\hat k)}\bar{\Gamma}^{(\hat\theta)} |
{\hat k}\in {\bf Z}_+^{n^2},  {\hat\theta}\in {\bf Z}_2^{2n}\}$
because of Lemmas \ref{Psi} and \ref{epsi}. It follows from
Lemma \ref{Psi} and equation (\ref{b})  that these elements
are linearly independent. Similarly we can show that
$\{\Gamma^{(\hat\theta)}F^{(\hat k)} |
\hat k\in {\bf Z}_+^{n^2}, \hat\theta\in {\bf Z}_2^{2n} \}$
forms a basis of $U_-$.
3) follows from 1) and 2).

\subsection{$\uq$ at roots of unity}
In this subsection we assume that $q$ is an $N$ - th primitive
root of unity. We define  $N'=\left\{\begin{array}{ll}
          N,&\mbox{$N$ odd},\\
          N/2, &\mbox{$N$ even.}
          \end{array}
   \right.$
Let $Z_q$ be the central algebra of the ${\bf Z}_2$ graded
algebra $\uq$ over the complex field $\bf C$, and let
$Z_0$ be the algebra generated by the following elements
\ba
(K_n^{\pm})^{N'},  &(K_i^{\pm})^N,\ \ i<n; &(E_{\beta_t})^N,
\ \ (F_{\beta_t})^N,\ \ t=1,2,...,n^2.
\na
It is  well known that\cite{concini} $Z_0$ is contained in the
central algebra of the maximal
even quantum subgroup $U_q(sp(2n)\oplus u(1))$. In fact, we also have
\begin{lemma}
$Z_q$ and $Z_0$ are as defined above. Then
\ba
Z_0\subset Z_q.
\na
\end{lemma}\label{center}
{\em Proof}: The proofs is exactly the same as Lemma 5 of reference \cite{I},
thus will not be repeated here.

$Z_0$ is a commutative algebra with no zero divisors. Following
\cite{concini} we define the quotient
field $Q(Z_0)$ of $Z_0$, and let $Q\uq=Q(Z_0)\x_{Z_0} \uq$.
Then $Q\uq$ is finite dimensional, with a basis
\ba
\{ \Gamma^{(\hat\theta)}F^{(\hat k)} K^{(\hat r)}E^{(\hat l)}
\bar{\Gamma}^{(\hat\theta')}
|{\hat k}, {\hat l} \in {\bf Z}_N^{n^2};
{\hat r}\in {\bf Z}_N^{n+1}, r_n\in {\bf Z}_{N'};
\hat\theta, \hat\theta' \in {\bf Z}_2^{2n}\}.
\na

\section{Finite dimensional representations }
\subsection{At generic $q$}
Given an irreducible module $V^{(0)}$ over the maximal even
quantum subgroup $U_q(sp(2n)\oplus U(1))$,
we construct a $\uq$ - module $\bar V$ over the field ${\bf C}[q, q^{-1}]$
in the following way. Impose the condition
\ba
e_0v=0,& \forall v\in V^{(0)},
\label{kac}
\na
and construct the ${\bf C}[q,q^{-1}]$ module
\ban
\bar{V}&=&\Phi\bigotimes{ _{{\bf C}[q,q^{-1}]}}V^{(0)}.
\nan
Given any element $a\in \uq$, and $\Gamma^{(\hat\theta)}\in \Phi$,
it follows from Theorem \ref{bpw} that
\ban
a\Gamma^{(\hat\theta)}=\sum c_{\hat\theta',\hat k,\hat r,\hat s,
\hat l,\hat\theta''}
\Gamma^{(\hat\theta')}F^{(\hat k)}K^{(\hat r)}H^{(\hat s)}
E^{(\hat l)}\bar{\Gamma}^{(\hat\theta'')}, &
c_{\hat\theta',\hat k,\hat r,\hat s,\hat l,\hat\theta''}
\in {\bf C}[q, q^{-1}].
\nan
We define the action of $a$ on $\bar {V}$ by
\ban
a(\Gamma^{(\hat\theta)}\x v)=
\sum c_{\hat\theta',\hat k,\hat r,\hat s,\hat l,0}
\Gamma^{(\hat\theta')}\x F^{(\hat k)} K^{(\hat r)}H^{(\hat s)}
E^{(\hat l)}v,
\nan
thus turning $\bar V$ into a $\uq$ - module. For simplicity, we will write
$\Gamma^{(\hat\theta)}\x v$ as  $\Gamma^{(\hat\theta)} v$ from here on.

Let $M$ be the maximal proper submodule contained in $\bar V$.  Setting
\ba
V=\bar{V}/M,
\na
we arrive at an irreducible $\uq$  module.
If $M=\{0\}$, we say that $V$ and the associated irrep of $\uq$
are typical, otherwise atypical.

The module $V$ has a ${\bf Z}_2$ gradation,
i.e., $V=V_0\oplus V_1$, with $V_0$ and $V_1$ respectively spanned by
$\{ \Gamma^{(\hat\theta)}v\in V | [\Gamma^{(\hat\theta)}]=0, v\in V^{(0)}\}$,
and
$\{ \Gamma^{(\hat\theta)}v\in V | [\Gamma^{(\hat\theta)}]=1, v\in V^{(0)}\}$.
This ${\bf Z}_2$ gradation is in consistency with that of $\uq$ itself, namely,
given any homogeneous element $a\in\uq$, we have
$a V_\epsilon \subset V_{\epsilon+[a] (mod 2)}, \ \epsilon=0,1$.
$V$ also has a natural $\bf Z$ gradation.   Let
\ban
deg(\Gamma^{(\hat\theta)})=\sum_{i=1}^n(\theta_i+\theta_{-i}), &
deg(\bar \Gamma^{(\hat\theta)})=-\sum_{i=1}^n(\theta_i+\theta_{-i}).
\nan
Define
$V^{(k)}$ to be the vector space spanned by
$\{\Gamma^{(\hat\theta)}v\in V | deg(\Gamma^{(\theta)})=k, v\in V^{(0)}\}$.
Then
\ban
V=\bigoplus_{k=0}^L V^{(k)},& L\leq 2n,
\nan
with
\ban
\Gamma^{(\hat\theta)} V^{(k)}\subset V^{(k+deg(\Gamma^{(\hat\theta)}))},
& V^{(l)}=\{0\}, \ \ \ \forall l>L, \\
\bar{\Gamma}^{(\hat\theta)} V^{(k)}\subset
V^{(k+deg({\bar\Gamma}^{(\hat\theta)}))},
& V^{(l)}=\{0\}, \ \ \ \forall l<0,
\nan
and each $V^{(k)}$ furnishes a $U_q(sp(2n)\oplus u(1))$ module,
\ban
aV^{(k)}\subset V^{(k)}, & \forall a\in U_q(sp(2n)\oplus u(1)).
\nan
In particular,
\begin{lemma}
1). $V^{(L)}$ is an irreducible $U_q(sp(2n)\oplus u(1))$  module;\\
2). $L$ is equal to $2n$ if and only if $V$ is typical; and in that case,
$V^{(2n)}$ and $V^{(0)}$ are isomorphic $U_q(sp(2n))$ modules.
\end{lemma}
{\em Proof}: 1) is required by the irreducibility of $V$.
To prove the first part of 2), we note that the necessity of
$L=2n$ is obvious. Let us assume that $L=2n$, i.e.,
$\Gamma V^{(0)}\not\subset M$,   but $V$ is atypical.
Then there must exist at least one nonvanishing vector  $u\in M$.
Now we can apply elements $\phi_{\pm i}$ to $u$ to obtain
another vector in $M$ of the form $\Gamma v$,
for some $v\in V^{(0)}$. Since $\Gamma$ commutes with all elements
of $U_q(sp(2n))$, the irreducibility of $V^{(0)}$ with respect to
$U_q(sp(2n))$ implies that
$\Gamma V^{(0)}\subset M$, which contradicts our assumption.
The second part of 2) follows from Lemma \ref{Gamma}.

Note that (\ref{kac}) is equivalent to $\psi_{\pm i} v=0$,
$\forall v\in V^{(0)}$. In any
given irreducible $\uq$ module, there always exists a subspace
obeying this condition.
Thus (\ref{kac}) does not impose any restrictions on
the irreducible module $V$, and the construction developed
above yields all irreps of the quantum supergroup $\uq$.

We will call $V$ a highest weight $\uq$ module if
there exists a unique vector $v^{\Lam}\in V$, with
$\Lam=\sum_{i=0}^n \lam_i \delta_i\in H^*$, $\lam_i\in{\bf C}$,
such that
\ba
e_iv^{\Lam}&=&0, \nonumber \\
\psi_{\pm i}v^{\Lam}&=&0,\ \ \ i=1,2,...,n, \nonumber \\
K_iv^{\Lam}&=&\epsilon_i q^{(\alpha_i,\ \Lam)}v^{\Lam},
\ \ \epsilon_i=\pm 1, \ \ \ i=0,1,...,n.
\label{highest}
\na
Observe that the sign factors $\epsilon_i$ may be eliminated
by the following isomorphism of $\uq$:
\ban
e_i\mapsto \epsilon_i^{-1}e_i,& f_i\mapsto f_i,& K_i\mapsto
\epsilon_i^{-1}K_i, \ \ \forall i.
\nan
Hereafter we will assume that to any irreducible $\uq$ module,
an appropriate isomorphism of
this kind has been employed to cast the last equation of
(\ref{highest}) into
\ban
K_iv^{\Lam}&=&q^{(\alpha_i,\ \Lam)}v^{\Lam}, \ \ \ i=0,1,...,n.
\nan

Since it is necessarily true that $v^{\Lam}\in V^{(0)}$,
$V$ is a highest weight module
if and only if  $V^{(0)}$ is of highest weight type.
To emphasize the role of the highest weight, we denote by
$V^{(0)}(\Lam)$ the $U_q(sp(2n)\oplus u(1))$ module $V^{(0)}$, and
introduce the new notations
$\bar{V}(\Lam)$, $V(\Lam)$, and  $M(\Lam)$, respectively,
for the $\bar V$, $V$ and $M$ constructed from $V^{(0)}$.

It immediately follows from our construction that the $\uq$
module $V(\Lam)$ is finite
dimensional if and only if the associated irreducible
$U_q(sp(2n)\oplus u(1))$ module $V^{(0)}(\Lam)$ is finite dimensional.
Since a finite dimensional irreducible $U_q(sp(2n)\oplus u(1))$
is uniquely characterized by its highest weight, so is the irreducible
$\uq$ module induced from it. Therefore,
\begin{proposition}
1). The irreducible highest weight $\uq$ module $V(\Lam)$ is
finite dimensional if and only if
\ba
{{2(\alpha_i, \Lam)}\over {(\alpha_i, \alpha_i)} } \in {\bf Z}_+,
&i=1,2,...,n;
\label{integer}
\na
2). A finite dimensional irreducible $\uq$ module $V(\Lam)$ is
uniquely determined by its  highest weight $\Lam$.
\end{proposition}
Note that $K_i$, $i=0,1,...,n$, are all diagonalizable on $V^{(0)}(\Lam)$.
Thus they are also
diagonalizable on the entire $\uq$ module $V(\Lam)$.
Define the weight space  $V_\omega\subset V(\Lam)$ to
be the vector space over ${\bf C}[q, q^{-1}]$  consisting of
all the vectors $v\in V(\Lam)$ satisfying $K_i v=q^{(\omega, \alpha_i)} v$.
Define $Sp_\Lam$
to be the set of all the distinct $\omega$'s such that
$\dimq V_\omega \ne 0$. Then
\ban
V(\Lam) = \oplus_{\omega\in Sp_\Lam} V_\omega.
\nan
By assigning the ordering $\delta_i>\delta_j>0$, $\forall i<j$,
we achieve a partial ordering of the elements of $H^*$ with
the same imaginary part.
Then it is clear that $\omega\le \Lam$, and
$\Lam-\omega=\sum_{i=0}^n m_i\alpha_i, \ \
m_i\in {\bf Z}_+$.

To gain further  understanding of the structures of $V(\Lam)$,
we now construct the highest weight vector of the irreducible
$U_q(sp(2n)\oplus u(1))$ module $V^{(L)}\subset
V(\Lam)$.
Consider the set of vectors
$\{v^{(0)},\  v^{(1)},..., v^{(L)}\}\subset V(\Lam)$  defined by
\ban
v^{(0)}&=&v^\Lam,\\
v^{(k)}&=&\phi_{\mu_k}v^{(k-1)}\ne 0, \\
\phi_{\nu}v^{(k-1)}&=&0, \ \ \mbox{if  $wt(\phi_\nu)>wt(\phi_{\mu_k})$},
\nan
where $\mu_k, \nu =\pm 1, \pm 2,...,\pm n$.
Since $\phi_i$, $i=1,2,..., n$, all $q$ - anticommute,
the existence of the vectors $\{v^{(0)},\  v^{(1)},..., v^{(l)}\}$
with $\mu_t>0$, $\forall t=1,2,...,l$, is guaranteed,
where $v^{(l)}$ is annihilated by all $\phi_i$.
Now if $v^{(l)}$ is also annihilated by all $\phi_{-i}$, then $l=L$;
otherwise there must exist
a $\phi_{\mu_{l+1}}$, $\mu_{l+1}<0$ which does not annihilated this vector,
but all $\phi_{-i}$ with $wt(\phi_{-i})>wt(\phi_{\mu_{l+1}})$ do.
We set $v^{(l+1)}=\phi_{\mu_{l+1}}v^{(l)}$.
Using Lemmas \ref{psipsi} we can easily see that
\ban
\phi_i v^{(l+1)}=0,& \forall i\\
\phi_{-i} v^{(l+1)}=0,& i>-\mu_{l+1}.
\nan
Continue this process we will eventually arrive at $v^{(L)}$.
It follows from the construction that
\begin{lemma}
All $v^{(k)}, \ \ k=0,1,..,L$,  are $U_q(sp(2n)\oplus u(1))$
highest weight vectors.
\end{lemma}
In particular, $v^{(L)}$ is  the highest weight vector of the irreducible
$U_q(sp(2n)\oplus u(1))$ module $V^{(L)}\subset V(\Lam)$.
If $V(\Lam)$ is typical, then $v^{(L)}=\Gamma v^\Lam$ can be raised
back to $v^\Lam$ by the action of $\bar\Gamma$.
Using the above Lemma and Lemma \ref{psiphi} we can compute
\ba
{\bar \Gamma}\Gamma v^\Lam &=& z(\Lam)v^\Lam, \nonumber\\
z(\Lam)&=&
\prod_{\gamma\in\Delta_1^+}{ {q^{(\Lam+\rho, \gamma)}-
q^{-(\Lam+\rho, \gamma)}}\over {q-q^{-1}}}.
\label{typical}
\na
Therefore,
\begin{proposition}
The irreducible highest weight $\uq$ module $V(\Lam)$ is
typical if and only if
\ba
z(\Lam)\ne 0,
\na
where $z(\Lam)$ is defined by (\ref{typical}).
\end{proposition}

On any irreducible highest weight $\uq$ module $V(\Lam)$ with
a real highest weight $\Lam$,
we introduce a sesquilinear form
$\langle .  | . \rangle : V(\Lam)\x V(\Lam)\rightarrow {\bf C}[q, q^{-1}]$,
which satisfies the following defining relations:  \\
1). If $v^\Lam\in V(\Lam)$ is the highest weight vector,
\ban
\langle v^{\Lam}|v^{\Lam}\rangle =1;
\nan
2).
\ban
\langle u| a v\rangle  =   \langle \omega(a)u| v\rangle,
& \forall a\in \uq, u, v\in V(\Lam),
\nan
where $\omega$ is the anti-automorphism defined before; \\
3).
\ban
\langle c_1u_1+c_2u_2| v\rangle& =& c_1^* \langle u_1| v\rangle +
c_2^* \langle u_2| v\rangle, \\
\langle v | c_1u_1+c_2u_2\rangle& =& c_1 \langle v | u_1\rangle +
c_2\langle v | u_2|\rangle,
\nan
where $c_1, c_2 \in {\bf C}[q, q^{-1}]$, $c_i^*=\omega(c_i)$,
$u_1, u_2, v \in V(\Lam)$.
Note that this form is well defined as long as the highest weight is real,
and has the standard property
$ \langle u|  v\rangle  =   (\langle u| v\rangle)^*,\ \ \
\forall  u, v\in V(\Lam)$.
Also,
\begin{lemma}
The form
$\langle .  | . \rangle : V(\Lam)\x V(\Lam)\rightarrow {\bf C}[q, q^{-1}]$
is nondegenerate.
\end{lemma}
{\em Proof}: The proof is rather straightforward, we nevertheless
present it here. Let $Ker \subset V(\Lam)$ be the maximal
subspace such that for any $k\in Ker$, $\langle v|k\rangle =0,
\ \ \  \forall v\in V(\Lam). $
Then $\langle v|a k\rangle = \langle \omega(a)v| k\rangle =0,
\ \ \  \forall a\in\uq,  \ \
v\in V(\Lam)$, i.e., $Ker$ is an invariant subspace.  Therefore
we must have $Ker=\{0\}$ as required by the irreducibility of $V(\Lam)$.

Now we compute the value of
$\langle v^{(L)} | v^{(L)} \rangle$,
which is nonvanishing as required by the the nondegeneracy of the form.
As $v^{(k)}$ are $U_q(sp(2n)\oplus u(1))$ maximal vectors, it follows
from Lemma \ref{psipsi} that $\psi_{\mu_k}v^{(k-1)}=0$. Thus
\ban
\langle v^{(k)} | v^{(k)} \rangle&=&\langle v^{(k-1)} |
{{\Pi_{\mu_k}- \Pi_{\mu_k}^{-1}}\over{q-q^{-1}}}  v^{(k-1)}\rangle.
\nan
Therefore,
\ba
\langle v^{(L)} | v^{(L)} \rangle
=\prod_{k=1}^L{ { q^{(\delta_0-\delta_{\mu_k}, \Lam + \sum_{t=1}^{k}
[\delta_0-\delta_{\mu_t}])} }
-q^{-(\delta_0-\delta_{\mu_k},\Lam+\sum_{t=1}^{k} [\delta_0-
\delta_{\mu_t}])}\over{q-q^{-1}} }.
\label{test1}
\na
Since $\langle v^{(L)} | v^{(L)} \rangle\ne 0$, we have
\ba
\prod_{k=1}^L (\delta_0-\delta_{\mu_k},\Lam + \sum_{t=1}^{k}
[\delta_0-\delta_{\mu_t}]) \ne 0.
\label{test2}
\na

Let ${\bf I}\subset{\bf C}[q, q^{-1}]$ be the ideal generated
by $q-1$. It is clear that
${\bf C}={\bf C}[q, q^{-1}]/{\bf I}$.  We define
$\tilde{V}(\Lam)= \{{\bf C}[q, q^{-1}]/{\bf I}\} \x V(\Lam)$, and
$\tilde{V}_\omega  = \{{\bf C}[q, q^{-1}]/{\bf I}\}\x V_\omega$
for any weight space
$V_\omega \subset V(\Lam)$. Then
\ban
\dimc \tilde{V}(\Lam)&=& \dimq V(\Lam), \\
{\tilde V}(\Lam)&=&\oplus_{\omega\in Sp_\Lam} {\tilde V}_\omega.
\nan
Denote by
$\tilde{e_i}$,  $\tilde{f_i}$,   $\tilde{h_i}$, and $1$  respectively
the endomorphisms on $\tilde{V}(\Lam)$
defined by the $V(\Lam)$ endomorphisms
$e_i$,  $f_i$,  ${{K_i-K_i^{-1}}\over{q_i-q_i^{-1}}}$, and $K_i^{\pm 1}$
through extension of scalars.  It can be proved that
\begin{lemma}
The $\tilde{e_i},\  \tilde{f_i},\   \tilde{h_i},\ \ i=0,1,...,n$,
satisfy the defining relations of the Lie superalgebra $C(n+1)$.
Thus $\tilde{V}(\Lam)$ furnishes a $U(C(n+1))$ module.
\end{lemma}
In particular,  $v^{\Lam}$ remains to be a highest weight vector in
$\tilde{V}(\Lam)$.   Repeatedly applying the $\tilde{f_i}$'s
to it  generates the entire $U(C(n+1)$ module $\tilde{V}(\Lam)$.
Therefore,
$\tilde{V}(\Lam)$
is indecomposible.  It immediately follows that
\begin{proposition}
The $U(C(n+1)$ module $\tilde{V}(\Lam)$ is typical and
irreducible if and only if the $\uq$ module $V(\Lam)$ is typical.
\end{proposition}

When the highest weight $\Lam$ is real, we denote the
restriction of the form
$\langle  . | .\rangle$ on $\tilde{V}(\Lam)$ by $\langle  . | .\rangle_0$,
which  maps
$\tilde{V}(\Lam)\x_{\bf C}\tilde{V}(\Lam)$ to $\bf C$.
Then  $\langle  . | .\rangle_0$ satisfies  similar properties as
$1) - 3)$. Furthermore,
\begin{proposition}\label{form}
The form
$\langle  . | .\rangle_0:  \tilde{V}(\Lam)\x_{\bf C}
\tilde{V}(\Lam)\rightarrow  \bf C$,
is nondegenerate.
\end{proposition}
{\em Proof}: Since the $U(C(n+1))$ module
$\tilde{V}(\Lam)$
is indecomposible, for every nonvanishing
$u\in \tilde{V}(\Lam)$
there exists at least one element $\tilde\phi\in U(C(n+1))$
which is a product of some $\tilde\phi_i$'s
(If $u\in \tilde{V}^{(L)}(\Lam)$, then $\tilde\phi=1$.),
such that the vector $v=\tilde\phi u \ne 0$,
and $v\in \tilde{V}^{(L)}(\Lam)$.
If the restriction of
$\langle  . | .\rangle_0$  on $\tilde{V}^{(L)}(\Lam)$ is nondegenerate,
then
$\langle v'|v\rangle_0$ does not vanish for some elements
$v'\in \tilde{V}^{(L)}(\Lam)$.  Now
\ban
\langle \tilde{\omega}(\tilde{\phi})v'|u\rangle_0 &=&
\langle v' | \tilde{\phi}u\rangle_0\ne 0,
\nan
where $\tilde{\omega}$ is the $q\rightarrow 1$ limit of
the anti automorphism $\omega$. Therefore the form
$\langle  . | .\rangle_0$  can not be degenerate on $\tilde{V}(\Lam)$.
The converse is also obviously true, thus we conclude that
$\langle  . | .\rangle_0$  is nondegenerate if and only
if it is nondegenerate on $\tilde{V}^{(L)}(\Lam)$.

It follows from the theorem of Lusztig and Rosso that
$\tilde{V}^{(L)}(\Lam)$
is an irreducible $sp(2n)\oplus u(1)$ module, with the highest
weight vector $\tilde{v}^{(L)}$ which is the $q\rightarrow 1$
limit of $v^{(L)}$.  Therefore
$\langle  . | .\rangle_0$  will be nondegenerate on
$\tilde{V}^{(L)}(\Lam)$  if
$\langle  \tilde{v}^{(L)}  |  \tilde{v}^{(L)} \rangle_0\ne 0.$
This is indeed that case, as it follows from (\ref{test1})
and (\ref{test2}) that
\ban
\langle  \tilde{v}^{(L)}  |  \tilde{v}^{(L)} \rangle_0
=\prod_{k=1}^L (\delta_0-\delta_{\mu_k}, \Lam+\sum_{t=1}^{k}
[\delta_0-\delta_{\mu_t}])
\ne 0.
\nan

The nondegeneracy of
$\langle  . | .\rangle_0$ implies that the indecomposible
$U(C(n+1))$ module $\tilde{V}(\Lam)$ is irreducible.
To see this, we note that if
$\tilde{V}(\Lam)$
was reducible, then there must exist at least one
$u \in \tilde{V}(\Lam)$ which could not be mapped to
the highest weight vector $v^\Lam$, or equivalently
\ban
\langle v^\Lam | \tilde{a}u\rangle_0 = 0,& \forall \tilde{a}
\in U(C(n+1)).
\nan
This would lead to
\ba
\langle \tilde{\omega}(\tilde{a})v^\Lam |u\rangle_0 = 0,
& \forall \tilde{a}\in U(C(n+1)).
\label{degeneracy}
\na
$\tilde{V}(\Lam)$ being an indecomposible $U(C(n+1))$ module,
every element of it
can be expressed as $\tilde{a} v^{\Lam}$, $\tilde{a}\in U(C(n+1))$.
Thus equation
(\ref{degeneracy}) would imply the degeneracy of $\langle  . | .\rangle_0$.

Combining the above discussion with Proposition \ref{form},
we arrive at  the following
\begin{theorem}\label{main}
Let $V(\Lam)$ be an irreducible $\uq$ module with an integrable dominant
highest weight $\Lam$(i.e., satisfying (\ref{integer})), and
$\tilde{V}(\Lam)$
be as defined before.  Then
$\tilde{V}(\Lam)$ is an irreducible
$U(C(n+1))$ module which has the same weight space decomposition
as $V(\Lam)$.
\end{theorem}

\noindent
{\bf Remarks}: \\
{\em 1). The form $\langle  . | .\rangle_0$ can be defined
independently of
$\langle  . | .\rangle$;\\
2). The proof of Theorem \ref{main} makes essential use of (\ref{test2}),
which can be proved
without resorting to the form $\langle  . | .\rangle$;\\
3). The forms  $\langle  . | .\rangle$ and $\langle  . | .\rangle_0$
are merely employed to  make the proof of Theorem \ref{main} more coherent;
they can be avoided entirely.} \\

Define the formal character of a finite dimensional irreducible
$\uq$ module $V(\Lam)$ by
\ban
ch_{V(\Lam)}=\sum_{\omega\in Sp_\Lam} \dimq V_\omega e^\omega.
\nan
Using Theorem \ref{main}, we obtain\cite{jeugt}
\begin{theorem}
Let $V(\Lam)$ be an irreducible $\uq$ module with an integrable
dominant highest weight $\Lam$.
Then
\ba
ch_{V(\Lam)}={{\prod_{\gamma\in\Delta_1^+(\Lam)}(e^{\gamma/2}
+e^{-\gamma/2})}
\over{\prod_{\alpha\in\Delta_0^+}(e^{\alpha/2}-e^{-\alpha/2})}}
\sum_{\sigma\in W} det(\sigma)e^{\sigma(\Lam+\rho)},
\na
where $W$ represents the Weyl group of the $sp(2n)\subset
C(n+1)$ subalgebra, and
\ban
\Delta_1^+(\Lam)
=\left\{\begin{array}{ll}
 \Delta_1^+,&\mbox{if $(\Lam+\rho, \gamma)\ne 0, \ \forall
\gamma\in\Delta_1^+$,}\\
 \Delta_1^+-\gamma_a,&\mbox{if $\exists \gamma_a\in\Delta_1^+$,
                            such that $(\Lam+\rho, \gamma_a)=0$. }
         \end{array}
 \right.
\nan
\end{theorem}
It should be noted that when $\Lam\in H^*$ is integrable dominant,
there can exist at most one odd root $\gamma_a\in\Delta_1^+$ rendering
$(\Lam+\rho, \gamma_a)= 0$, i.e., no two factors in the product
expression (\ref{typical}) of $z(\Lam)$ can vanish simultaneously.
Adopting the terminology of the representation theory of Lie superalgebras,
we say that a finite dimensional irrep of $\uq$ at generic $q$ is
either typical or singly atypical. We will
see in the next subsection that this is no longer true when
$q$ is a root of unity.

\subsection{At roots of unity}
The method developed in the last subsection for constructing
$\uq$ irreps works equally well
when $q$ is a root of unity. Because of Lemma \ref{form}, an irreducible
$U_q(sp(2n)\oplus u(1))$ module $V^{(0)}$ over $\bf C$
is necessarily finite dimensional, thus we conclude that
all irreps of $\uq$ at roots of unity are finite dimensional.

Properties of the irreducible $\uq$ module induced from
$V^{(0)}$
are completely determined by those of $V^{(0)}$, while
$V^{(0)}$
itself is uniquely characterized by a set of complex parameters
associated with the eigenvalues of the generators of $Z_0$.
We say that  $V$ is cyclic if the eigenvalues of all the
$(E_{\beta_t})^N$ and $(F_{\beta_t})^N$ are nonvanishing,
semicyclic if some are nonvanishing, and of highest weight type
otherwise.

Typicality of $V$ is defined in exactly the same way as in
the case with generic $q$.
$V$ is typical if and only if the eigenvalue  $z_{\mbox v}$
of $z$ defined by (\ref{z}) on $V^{(0)}$ is not zero. We have
\ban
dim_{\bf C}V&=&2^{2n}dim_{\bf C}V^{(0)},
\ \ \ \mbox{if $z_{\mbox v}\ne 0$}.
\nan

When $V$ is a highest weight module, there exists a
$v_0\in V$ such that
\ban
K_i v_0 &=& a_i v_0, \\
e_i v_0 &=& 0, \ \ \ i=0,1,...,n,
\nan
where $a_i$'s are complex parameters.
Then the eigenvalue $z_{\mbox v}$ of $z$ is given by
\ban
z_{\mbox v}&=&\prod_{\gamma\in\Delta_1^+}
{{\pi_\gamma q^{(\rho, \gamma)}-\pi_\gamma^{-1}q^{-(\rho, \gamma)}}
\over {q-q^{-1}}}, \\
\pi_{\delta_0-\delta_i}&=&\prod_{k=0}^{i-1} a_k, \\
\pi_{\delta_0+\delta_i}&=&\pi_{\delta_0 -\delta_n}\prod_{k=i}^{n} a_k.
\nan

Following the convention of the representation theory of Lie
superalgebras, we call an atypical $\uq$ module $V$ singly atypical
if only one factor in $z_{\mbox v}$ is zero, and multiply
atypical otherwise. There exist $a_i$ values rendering $V$
multiply atypical. Therefore,
\begin{lemma}
$\uq$ admits (semi)cyclic irreps and multiply atypical irreps
at roots of unity.
\end{lemma}
In contrast, all finite dimensional irreps of $\uq$ at generic
$q$ are of highest weight type,
and either typical or singly atypical.

\section{Irreducible Representations of $U_q(C(2))$}
In this section we construct all the highest weight irreps
of the quantum supergroup
$U_q(C(2))$
at generic $q$ and all the irreps at roots of unity.
For convenience, we change the notation from the general
case by letting
\ban
\psi_+=\psi_1,& \psi_-=\psi_{-1};\\
\phi_+=\phi_1,& \phi_-=\phi_{-1};\\
e=e_1,&         f=f_1.
\nan
When the deformation parameter $q$ is generic,
$U_q(C(2))$
has the following basis
\ban
\{(\phi_-)^{\theta_-}(\phi_+)^{\theta_+}
f^k K_0^m K_1^n e^l
(\psi_+)^{\theta'_+} (\psi_-)^{\theta'_-}
|k, l\in {\bf Z}_+,  \ m, n\in {\bf Z},\ \theta_{\pm},
\theta'_{\pm}\in\{0, 1\} \}.
\nan

Let $V(\Lam)$ be an irreducible $U_q(C(2))$ module with highest weight
$\Lam=\lam_0\delta_0+\lam_1\delta_1$, and maximal vector $v^\Lam$.
It is finite dimensional if and only if $\lam_1\in {\bf Z}_+$,
and in that case, $\Lam$
must satisfy one of the following three mutually exclusive conditions:
\ban
1). (\Lam+\rho, \gamma)\ne 0, \ \forall \gamma\in \Delta_1^+,&
2). (\Lam+\rho, \delta_0-\delta_1)=0,&
3). (\Lam+\rho, \delta_0+\delta_1)=0.
\nan
We explicitly construct $V(\Lam)$ for all the cases below:\\
1). $(\Lam+\rho, \gamma)\ne 0, \ \forall \gamma\in \Delta_1^+$:
\ban
V(\Lam)=\bigoplus_{\begin{array}{c}
                   \theta_\pm\in\{0,1\}\\
                   i\in\{0,1,...,\lam_1\}
                  \end{array} }
       {\bf C}[q, q^{-1}]\phi_-^{\theta_-}\phi_+^{\theta_+}f^i v^\Lam.
\nan
2). $(\Lam+\rho, \delta_0-\delta_1)=0$:
\ban
V(\Lam)=\left\{\begin{array}{ll}
 \bigoplus_{i\in\{0,1,...,\lam_1\}}{\bf C}[q, q^{-1}]f^i v^\Lam
 \bigoplus_{j\in\{0,1,...,\lam_1-1\}}{\bf C}[q, q^{-1}]f^j\phi_-v^\Lam,
                &\lam_1\ne 0,\\
               {\bf C}[q, q^{-1}]v^\Lam,& \lam_1=0.

		\end{array}
		 \right.
\nan
3). $(\Lam+\rho, \delta_0+\delta_1)=0$:
\ban
V(\Lam)=\bigoplus_{i\in\{0,1,...,\lam_1\}}{\bf C}[q, q^{-1}]f^i v^\Lam
\bigoplus_{j\in\{0,1,...,\lam_1+1\}}{\bf C}[q, q^{-1}]f^j\phi_+v^\Lam.
\nan

When $\lam_1\not\in{\bf Z}_+$, $V(\Lam)$ is infinite dimensional.
Then $V(\Lam)$ belongs to one of the following three cases:\\
1). $(\Lam+\rho, \gamma)\ne 0, \ \forall \gamma\in \Delta_1^+$:
\ban
V(\Lam)=\bigoplus_{\begin{array}{c}
                   \theta_\pm\in\{0,1\}\\
                   i\in{\bf Z}_+
                  \end{array} }
       {\bf C}[q, q^{-1}]\phi_-^{\theta_-}\phi_+^{\theta_+}f^i v^\Lam.
\nan
2). $(\Lam+\rho, \delta_0+\delta_1)=0$,  $(\Lam+\rho, \delta_0-\delta_1)\ne 0$:
\ban
V(\Lam)=\bigoplus_{\begin{array}{c}
                   \theta\in\{0,1\}\\
                   i\in{\bf Z}_+
                  \end{array} }
       {\bf C}[q, q^{-1}]f^i\phi_+^{\theta}v^\Lam.
\nan
3). $(\Lam+\rho, \delta_0-\delta_1)= 0$:
\ban
V(\Lam)=\bigoplus_{\begin{array}{c}
                   \theta\in\{0,1\}\\
                   i\in{\bf Z}_+
                  \end{array} }
       {\bf C}[q, q^{-1}]f^i\phi_-^{\theta}v^\Lam.
\nan

It is interesting to observe that in all the three cases with
$\lam_1\not\in{\bf Z}_+$, $V(\Lam)$ has finite dimensional weight
spaces. In the limit $q\rightarrow 1$, $V(\Lam)$ reduces to
an infinite dimensional irreducible
$U(C(2))$ module, which has the same weight space decomposition
as $V(\Lam)$ itself.

Now we assume that $q$ is an $N''$th  primitive root of unity.  Let
\ban
N'=\left\{\begin{array}{ll}
          N'',&\mbox{$N''$ odd,}\\
          N''/2, &\mbox{$N''$ even},
          \end{array}
    \right. &
N=\left\{\begin{array}{ll}
          N',&\mbox{$N'$ odd,}\\
          N'/2, &\mbox{$N'$ even}.
          \end{array}
    \right.
\nan
Then the following elements are all in the center of
$U_q(C(2))$:
\ban
(K_0^\pm)^{N''}, \ \ (K_1^\pm)^{N'}, \ \ e^{N'}, \ \ f^{N'},
\nan
provided $N'>1$.

We will call an irrep of
$U_q(C(2))$
a highest weight irrep if it possesses both a highest and lowest
weight vector, or equivalently,
\ban
det(e)=det(f)=0.
\nan
Such an irrep furnished by the irreducible module $V(a_0, a_1)$
is uniquely determined by
the two complex parameters $a_0,\ a_1$ defined in the following way:
Let $v_+$ be the highest weight vector of $V(a_0, a_1)$, then
\ba
K_0v_+=a_0 v_+,&   K_1v_+=a_1 v_+.
\na
We further define
\ban
d&=&\left\{\begin{array}{ll}
        i,&\mbox{if $a_1=\pm q^{-2(i-1)}$, with $N\ge i\ge 1$,}\\
        N,&\mbox{otherwise;}
        \end{array}
   \right.\\
{\tilde d}&=&\left\{\begin{array}{ll}
        d,&\mbox{if $a_1=\pm q^{-2(d-1)}$,}\\
        d-1,&\mbox{otherwise.}
        \end{array}
   \right.\\
V^{(0)}(a_0, a_1)&=&\bigoplus_{i=0}^d {\bf C}f^i v_+.
\nan
$V(a_0, a_1)$ can only belong to one of the following four cases: \\
1). $a_0-a_0^{-1}\ne 0$, $a_0a_1q^{-2}- a_0^{-1}a_1^{-1}q^2\ne 0$:
\ban
V(a_0, a_1)&=& V^{(0)}(a_0, a_1) \bigoplus V^{(1)}(a_0, a_1)
\bigoplus V^{(2)}(a_0, a_1),\\
V^{(1)}(a_0, a_1)&=&\bigoplus_{i=0}^{d-1} {\bf C}\phi_+f^i v_+\bigoplus
\{\bigoplus_{i=0}^{d-1} {\bf C}\phi_-f^i v_+\},\\
V^{(2)}(a_0, a_1)&=&\bigoplus_{i=0}^{d-1}{\bf C}\phi_+\phi_-f^i v_+.
\nan
2). $a_0-a_0^{-1}= 0$, $a_0a_1q^{-2}- a_0^{-1}a_1^{-1}q^2\ne 0$:
\ban
V(a_0, a_1)&=& V^{(0)}(a_0, a_1) \bigoplus V^{(1)}(a_0, a_1),\\
V^{(1)}(a_0, a_1)&=&\bigoplus_{i=0}^{{\tilde d}-1} {\bf C}\phi_-f^i v_+.
\nan
3). $a_0-a_0^{-1}\ne 0$, $a_0a_1q^{-2}- a_0^{-1}a_1^{-1}q^2= 0$:
\ban
V(a_0, a_1)&=& V^{(0)}(a_0, a_1) \bigoplus V^{(1)}(a_0, a_1),\\
V^{(1)}(a_0, a_1)&=&{\bf C}\phi_+ v_+
\bigoplus\{\bigoplus_{i=0}^{{\tilde d}-1} {\bf C}\phi_-f^i v_+\}.
\nan
4). $a_0-a_0^{-1}= a_0a_1q^{-2}- a_0^{-1}a_1^{-1}q^2= 0$:
\ban
V(a_0, a_1)&=& V^{(0)}(a_0, a_1) \bigoplus V^{(1)}(a_0, a_1),\\
V^{(1)}(a_0, a_1)&=&\bigoplus_{i=0}^{d-2} {\bf C}\phi_-f^i v_+.
\nan

Having explicitly constructed the highest weight irreps of
$U_q(C(2))$, we now consider the (semi)cyclic irreps.
We start with the simpler case that $N''$ is not divisible by $4$.
The (semi)cyclic irreducible module $V^{(0)}$ over the maximal
even subalgebra $U_q(sp(2)\oplus u(1))$
is $N$ dimensional, and labeled by $4$ parameters. Explicitly,
we have the following
basis $\{ v_i | i=0,1,...,N\}$ for $V^{(0)}$, with the actions
of the generators of    $U_q(sp(2)\oplus u(1))$ defined by
\ba
K_0 v_0 = a_0,& K_1 v_0 = a_1 v_0, \nonumber \\
e_1 v_0 = x v_{N-1}, & f_1 v_{N-1} = x' v_0, \nonumber \\
f_1 v_i = v_{i+1},& i=0, 1, ..., N-2,
\na
where the complex parameters $x$ and $x'$ do not vanish
simultaneously, and
\ba
x x'\ne { {(q^{2i} - q^{-2i})(a_1q^{2(i-1)}-a_1^{-1}q^{-2(i-1)})}
\over { q^2 -q^{-2}} },
& i=1,2,...,N-1.
\label{r1}
\na
For simplicity, we introduce the new parametrization
\ba
a_1=q^2 b b', &x={{u(b - b^{-1})}\over {q^2 - q^{-2}} },
&x'=-{{u^{-1}(b' - b'^{-1})}\over {q^2 - q^{-2}} },
\label{parameter}
\na
and also define
\ba
Q ={ {(a_0 b - a_0^{-1}b^{-1}) (a_0 b' - a_0^{-1}b'^{-1})}
\over {(q-q^{-1})^2} }.
\na
Denote by $V$ the irreducible (semi)cyclic $U_q(C(2))$ module
induced from $V^{(0)}$. Then \\
1). If $Q\ne 0$,
\ban
V&=& \bigoplus_{l=0}^2 V^{(l)},\\
V^{(1)}&=&\bigoplus _{i=0}^{N-1} \{{\bf C}\phi_+ v_i\oplus
{\bf C}\phi_- v_i\}, \\
V^{(2)}&=&\bigoplus _{i=0}^{N-1} {\bf C} \phi_-\phi_+v_i;
\nan
2). If $Q=0$, but either $x'\ne 0$ or $a_0 b - a_0^{-1} b^{-1} \ne 0$,
\ban
V&=&\bigoplus V^{(0)}\bigoplus V^{(1)},\\
V^{(1)}&=&\bigoplus _{i=0}^{N-1} {\bf C}\phi_- v_i;
\nan
3). If $Q=0$,  $x'= 0$, $a_0 b - a_0^{-1} b^{-1} =0$,
\ban
V&=&\bigoplus V^{(0)}\bigoplus V^{(1)},\\
V^{(1)}&=&\bigoplus _{i=0}^{N-2} {\bf C}\phi_- v_i.
\nan

When $N''$ is divisible by $4$, the (semi)cyclic irreducible
module $V^{(0)}$ over the maximal even subalgebra
$U_q(sp(2)\oplus u(1))$ is $2N$ dimensional,
with a basis $\{ v^{(\pm)}_i | i=0,1,...,N-1\}$ such that
\ba
e_1 v_0^{(\pm)} = \pm x v_{N-1}^{(\pm)}, & f_1 v_{N-1}^{(\pm)}
=\pm x' v_0^{(\pm)}, \nonumber \\
f_1 v_i^{(\pm)} = v_{i+1}^{(\pm)},& i=0, 1, ..., N-2,\nonumber \\
K_1 v_0^{(\pm)} = a_1 v_0^{(\pm)},& K_0 v_0^{(\pm )} = a_0 v^{(\mp)},
\na
where again $x$ and $x'$ do not vanish simultaneously,
and obey the constraint (\ref{r1}).
Note that from this basis we can always obtain a new one in
which $K_0^{(\pm 1)}$ are diagonal.
Let
\ban
V^{(0)}=V^{(0)}_+\bigoplus V^{(0)}_-,&  V^{(0)}_{\pm}
=\bigoplus_{i=0}^{N-1} {\bf C} v_i^{(\pm)}.
\nan
Then the irreducible $U_q(sp(2)\oplus u(1))$ module $V$ induced
from $V^{(0)}$ is given by\\
1). If $Q\ne 0$,
\ban
 V&=&\bigoplus_{l=0}^2 \{V_+^{(l)}\bigoplus V_-^{(l)}\},\\
V_\pm ^{(1)}&=&\bigoplus _{i=0}^{N-1} \{{\bf C}\phi_+ v^{(\pm)}_i
\oplus {\bf C}\phi_- v^{(\pm)}_i\}, \\
V_\pm^{(2)}&=&\bigoplus _{i=0}^{N-1} {\bf C} \phi_-\phi_+v^{(\pm)}_i;
\nan
2). If $Q=0$, but either $x'\ne 0$ or $a_0 b - a_0^{-1} b^{-1} \ne 0$,
\ban
V&=& \bigoplus_{\sigma=+, -} \{V_\sigma^{(0)}\bigoplus V_\sigma^{(1)}\},\\
V_\pm^{(1)}&=&\bigoplus _{i=0}^{N-1} {\bf C}\phi_- v^{(\pm)}_i;
\nan
3). If $Q=0$,  $x'= 0$, $a_0 b - a_0^{-1} b^{-1} =0$,
\ban
V&=& \bigoplus_{\sigma=+,-}\{ V_\sigma^{(0)}\bigoplus V_\sigma^{(1)}\},\\
V_{\pm}^{(1)}&=&\bigoplus _{i=0}^{N-2} {\bf C}\phi_- v^{(\pm)}_i.
\nan

\section{Conclusion}
We have presented a systematic treatment of the representation theory
of the quantum supergroup $\uq$.
The induced module construction developed here allows one to
construct the irreps of this quantum supergroup at arbitrary $q$.
Structures of the finite dimensional
irreps at generic $q$ have been investigated. In particular,
it has been shown that
every such irrep is a deformation of an irrep of the underlying
Lie superalgebra
$C(n+1)$. The character formula for the finite dimensional
irreps of $\uq$ is given.

We have also shown that when $q$ is a root of unity,
all irreps of $\uq$ are finite dimensional,
and (semi)cyclic irreps also exist.
The typicality criterion for highest weight irreps are given.
The structures of the typicals are understood, and the general
framework has also be set up for analyzing the structures
of the atypicals.  However, further investigation into this problem
will necessarily require detailed knowledge of the irreps of
the maximal even subalgebra $U_q(sp(2n)\oplus u(1))$ at roots
of unity, which is not available at present.

Reference \cite{I} and the present paper essentially complete
the representation theory
for the type I quantum supergroups at generic $q$.
With certain modifications,
techniques developed in these papers can also be generalized
to systematically
study the representation theory of the type II quantum supergroups.
Results will be reported in a forthcoming publication.

\end{document}